\documentclass{emulateapj}
\shorttitle{RELATIVISTIC FIREBALL}
\shortauthors{SUZUKI\&SHIGEYAMA}
\begin{document}
\title{DYNAMICAL EVOLUTION OF AN ULTRA-RELATIVISTIC FIREBALL COLLIDING WITH A FREELY EXPANDING GAS}
\author{AKIHIRO SUZUKI\altaffilmark{1} and TOSHIKAZU SHIGEYAMA\altaffilmark{2}}
\altaffiltext{1}{Department of Astronomy, Kyoto University, Kitashirakawa-Oiwake-cho, Sakyo-ku, Kyoto, 606-8502, Japan}
\altaffiltext{2}{Research Center for the Early Universe, School of Science, University of Tokyo, Bunkyo-ku, Tokyo, 113-0033, Japan.}
\begin{abstract}
We investigate the hydrodynamical evolution of an ultra-relativistic fireball colliding with a freely expanding gas. 
The hydrodynamical interaction of the fireball and the gas results in the formation of a geometrically thin shell. 
We study the dynamical evolution of the shell by an analytical way and perform a numerical simulation equipped with an adaptive mesh refinement to investigate the internal structure of the shell. 
The shocked gas can give rise to bright emission in the X-ray and gamma-ray energy range. 
We propose that the breakout emission from the forward shock and the photospheric emission from the reverse-shocked fireball contribute to early gamma-ray emission from gamma-ray bursts. 
\end{abstract}

\section{INTRODUCTION\label{intro}}
Hydrodynamics of relativistic outflows is of crucial importance in many astrophysical phenomena. Some high-energy astrophysical phenomena found in X-ray and gamma-ray observations can naturally be explained by introducing relativistic outflows because the observed energy of a photon emitted from a particle moving at a highly relativistic speed toward an observer could become much higher than that in the rest frame of the particle.  For example, emission from an expanding hot gas at relativistic speeds in spherical symmetry, which is often called a relativistic fireball, is a key ingredient to understand the dynamics of gas in extremely energetic explosive phenomena.

Gamma-ray bursts are one of the explosive phenomena. 
They are characterized by sudden appearance of a bright gamma-ray point source on the celestial sphere \citep[see, e.g.,][for review]{2006RPPh...69.2259M} and thought to originate from stellar explosions at cosmological distances. 
A collimated jet launched from a compact object at relativistic speeds is needed to account for the bright gamma-ray emission. 
Although the mechanism to produce highly energetic photons is still in debate, some mechanisms to dissipate a fraction of the kinetic energy of the flow are indispensable to account for the total energy of the gamma-ray emission. 

It has been recognized that spectra of GRBs are well fitted by a broken power-law, i.e., the so-called Band function \citep{1993ApJ...413..281B}. 
The most widely discussed model to explain spectral and temporal features of the prompt gamma-ray emission from GRBs is the internal shock model, in which shocks propagating in an ultra-relativistic jet dissipate a part of the kinetic energy of the jet and produce non-thermal particles capable of emitting highly energetic photons. 

Recent observations of GRBs by the BATSE instrument on the {\it Compton Gamma Ray Observatory (CGRO)} and the {\it Fermi} satellite revealed that spectra of some GRBs consist of a component well fitted by a Planck function in addition to non-thermal components \citep{2004ApJ...614..827R,2005ApJ...625L..95R,2006ApJ...652.1400R,2009ApJ...702.1211R,2010ApJ...709L.172R,2011ApJ...727L..33G,2012ApJ...757L..31A}, suggesting the presence of photospheres in ultra-relativistic jets.
The photospheric emission from an ultra-relativistic jet is thought to play important roles in producing the prompt gamma-ray emission. 
From a theoretical point of view, contributions of the photospheric emission to GRB spectra have been widely discussed \citep{1986ApJ...308L..47G,1986ApJ...308L..43P,1994MNRAS.270..480T}. 
Since the discovery of black-body components in GRB spectra, a great attention has been paid to  photospheric emission models for the prompt gamma-ray emission and investigations into the mechanism to modify a Planck function into a broken power-law by some dissipative process have been put forward \citep[e.g.,][]{2006A&A...457..763G,2007A&A...469....1G,2010MNRAS.407.1033B,2010ApJ...725.1137L}. 

The evolution of an ultra-relativistic jet and the interaction with the ambient gas are of great importance in determining the spectral and temporal features of the gamma-ray emission. 
Especially, shock waves are an efficient and ubiquitous process to convert the kinetic energy of a flow into the internal energy. 
Therefore, the dynamics of relativistic shock waves in various situations has been considered. 
The Blandford-McKee solution \citep{1976PhFl...19.1130B}, the relativistic extension of the Sedov-Taylor point explosion problem, is one of the well-known examples. 
The dynamical evolution of a relativistic shock in a stellar atmosphere whose density profile is described as a power of the distance from the surface has been studied by self-similar approach \citep{2005ApJ...627..310N,2006ApJ...643..416P} and numerical simulations \citep{2007ApJ...657..860K}. 
The interaction of freely expanding ejecta moving at relativistic speeds with an ambient gas is also important because it could also give rise to bright emission. 
A self-similar solution describing the hydrodynamical interaction was discovered by \cite{2006ApJ...645..431N}.

\begin{figure}[tbp]
\begin{center}
\includegraphics[scale=0.4]{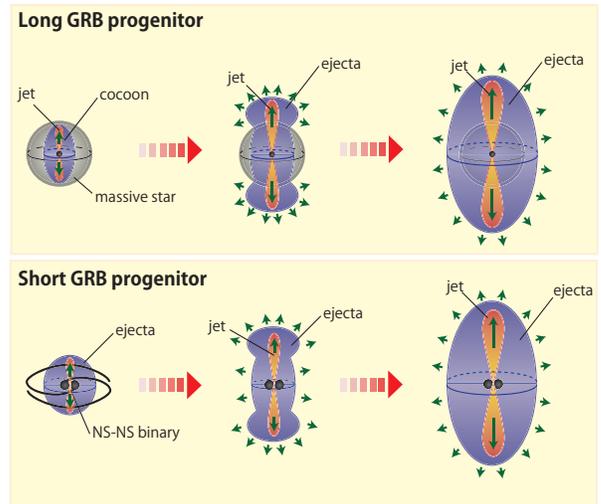}
\caption{Schematic views of the plausible scenarios for the long (upper panel) and short (lower panel) GRB progenitor systems.  
In both scenarios, the hydrodynamical interaction of an ultra-relativistic fireball with freely expanding ejecta could be realized.}
\label{fig:schematic}
\end{center}
\end{figure}

As the progenitor of long-duration GRBs, the gravitational collapse of the core of a massive star in its final evolutionary stage is the most plausible scenario because of the so-called GRB-supernova connection \citep[see, e.g.,][for review]{2006ARA&A..44..507W}. 
The scenario is schematically illustrated in the upper panel of Figure \ref{fig:schematic}. 
In this scenario, an ultra-relativistic jet emanating a massive star is responsible for the prompt gamma-ray emission. 
The injection and propagation of a jet in a massive star have been extensively studied both analytically and numerically. 
For example, \cite{2011ApJ...740..100B} analytically dealt with the propagation of a jet in an ambient gas and obtained a criterion for the collimation of the jet. 
There are a number of numerical studies on the dynamical evolution of GRB jets \citep{2000ApJ...531L.119A,2003ApJ...586..356Z,2006ApJ...651..960M,2007ApJ...665..569M}. 
These studies clarified that the penetration of the jet into the stellar mantle stratified on the jet results in the formation of a hot cocoon surrounding the jet (see the upper panel of Figure \ref{fig:schematic}). 
Since the cocoon is created by the hydrodynamical interaction of the jet with the star, the gas in the cocoon moves at subsonic speeds. 
Therefore, the gas starts expanding into the interstellar space in a nearly spherical manner after the cocoon emerges from the stellar surface. 
Hereafter, we call the expanding gas as "ejecta". 
When the jet injection continues after the emergence, the injected jet can leave the star through a hole created by the jet penetration and propagate almost freely following the ejecta. 
Thus, freely expanding ejecta pushed by an ultra-relativistic fireball naturally realize in this scenario. 
In addition, one can expect that the photospheric emission from the ejecta powered by the jet would contribute to the early gamma-ray emission from GRBs when the density of the ejecta is sufficiently high. 

On the other hand, for short GRBs, the merger of double neutron stars (NSs) in a closed binary system is a promising scenario \citep{1992ApJ...395L..83N}, which is schematically illustrated in the lower panel of Figure \ref{fig:schematic}. 
Numerical relativity is a powerful tool to investigate the dynamics of NS-NS mergers and resultant gravitational wave signals, which may be detected by next-generation gravitational wave detectors, such as, advanced LIGO, advanced VIRGO, and KAGRA \citep[see,][for review]{2010CQGra..27k4002D,2012LRR....15....8F}. 
Recent numerical simulations of NS-NS mergers based on numerical general relativity revealed that materials having been bound in the gravitational potential of the binary could be accelerated to the escape velocity of the system due to the heating by the shock generated from the impact of the merger \citep{2013PhRvD..87b4001H}. 
The ejected gas travels at mildly relativistic speeds in a nearly spherical manner. 
Then, it gradually approaches to free expansion. 
An ultra-relativistic jet responsible for the prompt gamma-ray emission is expected to be launched shortly after the merger. 
Thus, the launched jet propagates in the ejected material, resulting in the hydrodynamical interaction of an ultra-relativistic fireball with a freely expanding gas \citep{2014ApJ...784L..28N,2014arXiv1404.0383M}.

In other words, the hydrodynamical interaction of an ultra-relativistic fireball with an expanding gas naturally realizes in potential long and short GRB progenitors. 
Then, in this paper, we consider an ultra-relativistic fireball following expanding ejecta and investigate their hydrodynamical interaction in spherical symmetry. 
In Section \ref{evolution}, the dynamical evolution of the gas is studied in approximate and numerical ways. 
The propagation of the shocks forming as results of the hydrodynamical interaction between the gases is investigated in detail in Section \ref{shock_waves}. 
Then, we discuss possible processes to produce high-energy emission in Section \ref{sec:emission}. 
We conclude this paper in Section \ref{sec:conclusions}. 
In the following, we use the unit $c=1$ where $c$ denotes the speed of light. 

\section{DYNAMICAL EVOLUTION OF EJECTA}\label{evolution}
In this section, we consider the dynamical evolution of ejecta expanding into the interstellar space and being followed by an ultra-relativistic fireball in spherical symmetry. 
Thus, hydrodynamical variables are functions of the time $t$ and the radial coordinate $r$. 
We denote the radial velocity, the density, and the pressure of the gas by, $\beta(r,t)$, $\rho(t,r)$, and $p(t,r)$. 
The dynamical evolution of these variables is governed by the following hydrodynamical equations in spherical symmetry,
\begin{equation}
\frac{\partial (\rho\Gamma)}{\partial t}+\frac{\partial (r^2\rho\Gamma\beta)}{\partial (r^3/3)}=0,
\label{eq:continuity}
\end{equation}
\begin{equation}
\frac{\partial (\rho h\Gamma^2\beta)}{\partial t}+\frac{\partial [r^2(\rho h\Gamma^2\beta^2)+p]}{\partial (r^3/3)}=\frac{2p}{r},
\label{eq:momentum}
\end{equation}
and
\begin{equation}
\frac{\partial (\rho h \Gamma^2-p)}{\partial t}+\frac{\partial (r^2\rho h \Gamma^2\beta)}{\partial (r^3/3)}=0,
\label{eq:energy}
\end{equation}
where the Lorentz factor $\Gamma$ is expressed in terms of the velocity as,
\begin{equation}
\Gamma=\frac{1}{\sqrt{1-\beta^2}},
\end{equation}
and the specific enthalpy $h$ for an ideal gas with an adiabatic index $\gamma$ is given by
\begin{equation}
h=1+\frac{\gamma}{\gamma-1}\frac{p}{\rho}.
\label{eq:enthalpy}
\end{equation}
In this work, we assume that the gas is radiation-dominated and thus the adiabatic index is fixed to be $4/3$.

\subsection{Fireball Solution}
At first, we consider the profiles of the physical variables in the fireball. 
The relativistic fireball solution \citep{1993ApJ...415..181M,1993MNRAS.263..861P,1999ApJ...513..669K} is a well-known solution for the special relativistic hydrodynamical equations in spherical symmetry. 
A great attention has been paid to this solution to reveal the nature of GRB jets. 
In the following, we briefly review the solution. 

We consider a gas continuously injected from $r=R_\mathrm{in}$ at constant mass and energy injection rates, $\dot{M}$ and $L$. 
It is convenient to introduce a non-dimensional parameter, $\eta$, which gives the ratio of the energy injection rate to the mass injection rate,  
\begin{equation}
\eta=\frac{L}{\dot{M}}. 
\end{equation}

We assume that the flow is ultra-relativistic, $\beta\sim 1$. 
Thus, the balance of the mass and the energy fluxes, Equations (\ref{eq:continuity}) and (\ref{eq:energy}), at $r$ yields
\begin{equation}
4\pi r^2\rho\Gamma=\dot{M},
\end{equation}
and
\begin{equation}
4\pi r^2\rho h\Gamma^2=L.
\end{equation}
Furthermore, the gas is adiabatic along the streamline, 
\begin{equation}
p\propto \rho^{4/3}.
\end{equation}

When the internal energy of the gas dominates over its rest mass energy,
 $p\gg \rho$, 
the gas expands by converting its internal energy into the kinetic one. 
In this case, the dependence of the Lorentz factor, the density, and the pressure on the radial coordinate is found to be,
\begin{equation}
\Gamma\propto r,\ \ \ 
\rho\propto r^{-3},\ \ \ 
p\propto r^{-4}. 
\label{eq:fireball1}
\end{equation}
On the other hand, for a gas with the rest mass energy much larger than the internal one, $p\ll \rho$,
 the radial profiles of the variables are as follows,
\begin{equation}
\Gamma=\eta,\ \ \ 
\rho\propto r^{-2},\ \ \ 
p\propto r^{-8/3},
\label{eq:fireball2}
\end{equation}
which is identical with a wind solution with constant velocity and mass-loss rate. 

When a gas with the internal energy much larger than the rest mass energy is released in a small region, the gas initially expands according to Equations (\ref{eq:fireball1}). 
Then the kinetic energy eventually dominates over the internal one and the gas finally reaches to the state well described by Equations (\ref{eq:fireball2}). 

\subsection{Thin Shell Approximation\label{thin_shell}}
When the ejecta pushed by an ultra-relativistic fireball are slower than the fireball and the pressure at the interface between the ejecta and the fireball is sufficiently small, the forward and reverse shocks form at the interface and the swept gas forms a geometrically thin shell. 
We call the resultant shocked gas as the "shell" hereafter. 
Then, before going to numerical calculations, we derive the dependence of physical variables of the shell on the time $t$ using a thin shell approximation. 

We denote the mass of the shell by $M_\mathrm{shell}$. 
The shell is accelerated by the pressure gradient inside the shell. 
When we denote the pressure of the post-shock gas of the forward and the reverse shocks by $p_\mathrm{fs}$ and $p_\mathrm{rs}$, the equation of motion of the shell at a position $r=R_\mathrm{shell}$, which governs the time dependence of the Lorentz factor $\Gamma_\mathrm{shell}$ of the shell, is expressed as follows,
\begin{equation}
\frac{d(M_\mathrm{shell}\Gamma_\mathrm{shell}\beta)}{dt}=4\pi R_\mathrm{sh}^2(p_\mathrm{rs}-p_\mathrm{fs}).
\end{equation}
Here, we assume the Lorentz factor $\Gamma_\mathrm{shell}$ to be much larger than unity and consider the following limit, $\beta\sim 1$. 
Thus, the position $R_\mathrm{shell}$ of the shell is proportional to the time $t$, $R_\mathrm{shell}\propto t$. 
Furthermore, the post-shock pressure $p_\mathrm{fs}$ at the forward shock is assumed to be much smaller than that $p_\mathrm{rs}$ at the reverse shock, $p_\mathrm{fs}\ll p_\mathrm{rs}$. 
Then, the equation of motion can be approximated as follows,
\begin{equation}
\frac{d(M_\mathrm{shell}\Gamma_\mathrm{shell})}{dt}\propto t^2p_\mathrm{rs}.
\end{equation}
The dependence of the post-shock pressure $p_\mathrm{rs}$ is determined by the shock jump condition at the reverse shock. 
The pressure is proportional to the product of the density of the fireball and the square of the ratio of the Lorentz factors, $\Gamma_\mathrm{f}$ and $\Gamma_\mathrm{shell}$, of the fireball and the shell under the strong shock approximation (see Appendix \ref{jump_condition}, for the derivation of the jump condition at the reverse shock),
\begin{equation}
p_\mathrm{rs}\propto \rho_\mathrm{f}\frac{\eta^2}{\Gamma_\mathrm{shell}^2}\propto t^{-2}\Gamma_\mathrm{shell}^{-2},
\end{equation}
where we denote the pre-shock density at the reverse shock by $\rho_\mathrm{f}$, which is proportional to the inverse square of the time, $\rho_\mathrm{f}\propto t^{-2}$, according to Equation (\ref{eq:fireball2}). 
We regard the Lorentz factor $\eta$ of the fireball as a constant, because the fireball approaches to a steady wind solution at a large distance. 
The equation of motion is finally written as,
\begin{equation}
\frac{d(M_\mathrm{shell}\Gamma_\mathrm{shell})}{dt}\propto \Gamma_\mathrm{shell}^{-2}.
\end{equation}
This equation can be solved once the dependence of the mass $M_\mathrm{shell}$ on the time $t$ is determined. 
We consider cases where the mass is proportional to a power of the time $t$, $M_\mathrm{shell}\propto t^{\alpha}$. 
The exponent $\alpha$ cannot be negative as long as the system evolves in spherical symmetry, because the mass of the gas swept by the reverse and the forward shocks would increase with time. 
For example, one may assume that the mass of the shell is dominated by that of the ejecta and most of the ejecta have been swept by the forward shock. 
In this case, the mass hardly changes with time, $\alpha=0$.

Under this assumption, the temporal behavior of the Lorentz factor turns out to be
\begin{equation}
\Gamma_\mathrm{shell}\propto t^{(1-\alpha)/3}.
\end{equation}
We also obtain the time dependence of the pressure at the reverse shock as,
\begin{equation}
p_\mathrm{rs}\propto t^{2(\alpha-4)/3}.
\end{equation}

\begin{figure}[tbp]
\begin{center}
\includegraphics[scale=0.45]{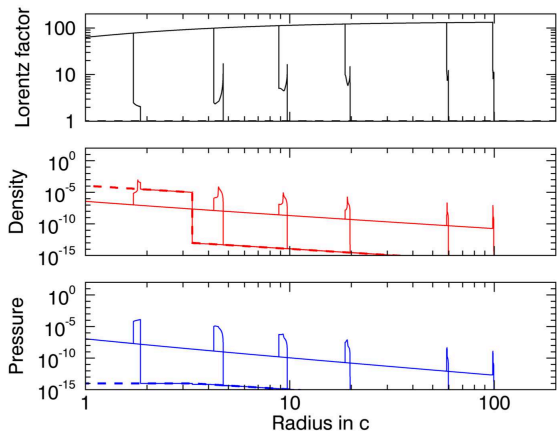}
\caption{Results of a numerical calculation of a fireball colliding with a stationary gas. 
In the panels, the radial profiles of the Lorentz factor (top), the density (middle), and the pressure (bottom) at $t=2.0$, $5.0$, $10.0$, $20.0$, $60.0$, and $100.0$ s are shown as thin solid lines. 
The initial profiles for these variables are also plotted as thick dashed lines. }
\label{fig:evolution}
\end{center}
\end{figure}

\subsection{Numerical Simulations\label{numerical}}
We have demonstrated that a geometrically thin shell pushed by an ultra-relativistic fireball accelerates due to the pressure gradient inside the shell under the thin shell approximation. 
In order to obtain the profiles of hydrodynamical variables in the shell, we have to numerically integrate the hydrodynamical equations (\ref{eq:continuity}) - (\ref{eq:enthalpy}). 
Numerical methods including the adaptive mesh refinement used in this work are briefly described in Appendix \ref{numerical_techniques}. 

\subsubsection{Initial Conditions and Fireball Injection}
Numerical calculations are performed in a spherical coordinate system from $r=R_\mathrm{in}$ to $r=R_\mathrm{out}$. 
In this simulation, an expanding gas is created by injecting a part of the energy from the fireball into a static gas, imitating situations expected to realize in long GRB progenitors. 
Initially, a static gas with the density inversely proportional to the square of the radius is distributed from $r=R_\mathrm{in}$ to $r=R_\ast$, and surrounded by a dilute gas with a steady wind profile, which is referred to as the circum-stellar medium (CSM),
\begin{equation}
\rho(r,0)=
\left\{\begin{array}{ccl}
\rho_\ast\left(\frac{r}{R_\ast}\right)^{-2}&\mathrm{for}&r\leq R_\ast,\\
\rho_\mathrm{csm}\left(\frac{r}{R_\ast}\right)^{-2}&\mathrm{for}&R_\ast<r.
\end{array}
\right.
\end{equation}
The gas is assumed to be cold. 
However, zero pressure cannot be treated in the numerical code used in this work. 
Thus, we use sufficiently small, but non-zero values for the pressure of the CSM. 
The following pressure profile is employed so that the pressure of the gas does not interrupt the propagation of shocks resulting from the impact of the fireball injection,
\begin{equation}
p(r,0)=
\left\{\begin{array}{ccl}
0.1\rho_\mathrm{csm}&\mathrm{for}&r\leq R_\ast,\\
0.1\rho_\mathrm{csm}\left(\frac{r}{R_\ast}\right)^{-2}&\mathrm{for}&R_\ast<r.
\end{array}
\right.
\end{equation}

The boundary condition at $r=R_\mathrm{in}$ must be specified to launch a fireball. 
For a given set of the initial Lorentz factor $\Gamma_\mathrm{in}$, the kinetic luminosity $L$, and the parameter $\eta$, the following conditions for the velocity, the density, and the pressure are imposed at $r=R_\mathrm{in}$,
\begin{equation}
\beta_\mathrm{in}=\sqrt{1-\frac{1}{\Gamma_\mathrm{in}^2}},
\end{equation}
\begin{equation}
\rho_\mathrm{in}=\frac{L}{4\pi \eta R_\mathrm{in}^2\Gamma_\mathrm{in}},
\end{equation}
and 
\begin{equation}
p_\mathrm{in}=\frac{\gamma-1}{\gamma}\left(\frac{\eta}{\Gamma_\mathrm{in}}-1\right)\rho_\mathrm{in}.
\end{equation}
The pressure at the inner boundary $p_\mathrm{in}$ can also be expressed in terms of the specific internal energy $\epsilon_\mathrm{in}$,
\begin{equation}
p_\mathrm{in}=(\gamma-1)\rho_\mathrm{in}\epsilon_\mathrm{in}.
\end{equation}

\subsubsection{Results}
We carry out a simulation with the following parameters for the initial configuration of the gas, $R_\mathrm{in}=10^9$ cm, $R_\ast=10^{11}$ cm, $R_\mathrm{out}=6\times 10^{12}$ cm, $\rho_\ast=10^{-3}$ g cm$^{-3}$, and $\rho_\mathrm{csm}=10^{-9}$ g cm$^{-3}$. 
The numerical domain is covered by $2048$ cells with the refinement level $l=0$ and the maximum refinement level is set to $l_\mathrm{max}=10$. 

\begin{figure*}[tbp]
\begin{center}
\includegraphics[scale=0.9]{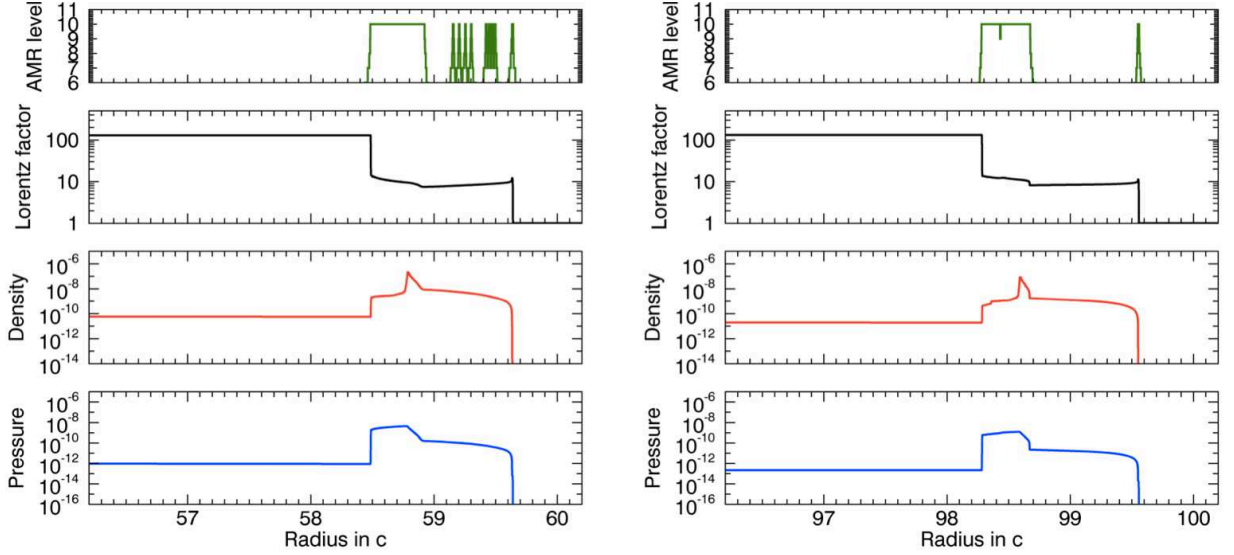}
\caption{Enlarged view of the radial profiles of the Lorentz factor, the density, and the pressure around $r=60$ at $t=60$ s (left panel) and $r=100$ at $t=100$ s (right panel). 
The AMR refinement level as a function of the radial coordinate $r$ is also plotted in the top panel.}
\label{fig:profile100}
\end{center}
\end{figure*}

Results of a simulation with an energy injection rate $L=10^{51}$ erg s$^{-1}$, an initial Lorentz factor $\Gamma_\mathrm{in}=5$, and $\eta=138$ (or equivalently $\epsilon_\mathrm{in}=20$), are presented in Figures \ref{fig:evolution} and \ref{fig:profile100}. 
Figure \ref{fig:evolution} shows the snapshots of the radial profiles of the Lorentz factor, the density, and the pressure at $t=2.0$, $5.0$, $10.0$, $20.0$, $60.0$, and $100.0$ s. 
At first, the fireball injected from the inner boundary generates a forward shock propagating in the static gas at $r<R_\ast$ in earlier stages of the dynamical evolution.  
After the forward shock finishes sweeping the gas, the shocked gas starts expanding into the surrounding medium, which corresponds to the "ejecta". 
The expanding gas is followed by the fireball, resulting in a geometrically thin shell connecting to the unshocked fireball through the reverse shock. 
The radial profiles of the hydrodynamical variables of the fireball are in good agreement with the relations given in the previous section, Equations (\ref{eq:fireball2}), after the Lorentz factor saturates. 

\subsubsection{Structure of the Shell}\label{sec:structure_of_shell}
The ejecta expand adiabatically after the passage of the forward shock. 
Thus, the pressure of the ejecta evolves as $\propto t^{-4}$, which decreases faster than that of the post-shock pressure of the reverse shock in the fireball. 
The difference between the pressure of the preceding ejecta and that of the reverse-shocked fireball gets larger as time elapses, resulting in the formation of a shock propagating into the ejecta. 

Finally, the system is composed of the following layers from the inner boundary to the outer edge of the ejecta, 
(1) unshocked fireball, 
(2) reverse-shocked fireball, 
(3) forward-shocked ejecta, 
and (4) unshocked ejecta.
Figure \ref{fig:profile100} presents the radial profiles of the Lorentz factor, the density and the pressure of the shell at $t=60$ and $100$ s. 
The forward shock eventually develops after the pressure of the preceding ejecta becomes sufficiently smaller than that of the reverse-shocked fireball. 
In the left panel of Figure \ref{fig:profile100}, the forward shock is not clearly recognized yet. 
On the other hand, the forward shock and the layers described above are clearly seen in the right panel of Figure \ref{fig:profile100}. 

On the other hand, at the interface between the ejecta and the ambient stationary gas, a pair of waves, rarefaction-forward shock or reverse shock-forward shock, is expected to develop, depending on the pressure of the ejecta and the density of the ambient gas. 
In the simulation, the density of the ambient gas is set to a sufficiently small value so that the ejecta are hardly affected by the interaction with the ambient gas. 
As a consequence, the resultant ejecta travel almost freely as seen in Figure \ref{fig:profile100}. 
In the following sections, we focus on the reverse and forward shocks at the fireball-ejecta interface and do not consider the waves expected to develop at the ejecta-ambient gas interface for the sake of simplicity.

\section{PROPAGATION OF SHOCK WAVES}\label{shock_waves}
In the previous section, we have demonstrated that a shock naturally forms in the expanding ejecta due to the fireball-ejecta interaction. 
In the numerical simulation, we have created a freely expanding gas by injecting an energy into a static gas at the same rate as the fireball. 
However, in long and short GRB progenitors, ejecta are expected to have various density and Lorentz factors. 
In this section, we generally discuss the propagation of the forward and reverse shocks forming at the interface between the fireball and the ejecta. 

\subsection{Forward Shock}
At first, we consider the temporal evolution of the unshocked ejecta. 
For ejecta with the maximum Lorentz factor $\Gamma_\mathrm{max}$, we assume the following profiles for the density and pressure,
\begin{equation}
\rho=\rho_0 \left(\frac{t}{t_0}\right)^{-3}\left(\frac{\Gamma}{\Gamma_\mathrm{max}}\right)^{-n},
\label{eq:adiabatic_rho}
\end{equation}
and
\begin{equation}
p=p_0 \left(\frac{t}{t_0}\right)^{-4}\left(\frac{\Gamma}{\Gamma_\mathrm{max}}\right)^{-s},
\label{eq:adiabatic_pre}
\end{equation}
with the Lorentz factor given by
\begin{equation}
\Gamma=\frac{1}{\sqrt{1-(r/t)^2}}.
\label{eq:free_gamma}
\end{equation}
This is a solution for the hydrodynamical equations (\ref{eq:continuity}) - (\ref{eq:enthalpy}) with $\beta=r/t$ and $p\ll \rho$.

\subsubsection{Passage of a Strong Shock}
We consider the temporal evolution of a shock with the Lorentz factor expressed by a power-law function of the time $t$,
\begin{equation}
\Gamma_\mathrm{fs}^2=A t^{-m}.
\label{eq:gamma_fs}
\end{equation}
Integration of the shock velocity, $\beta_\mathrm{fs}\simeq 1-1/(2\Gamma_\mathrm{fs}^2)$, with respective to the time from $0$ to $t$ gives the position $R_\mathrm{fs}$ of the forward shock at $t$,
\begin{equation}
R_\mathrm{fs}=t\left[1-\frac{1}{2(m+1)\Gamma_\mathrm{fs}^2}\right].
\label{eq:shock_position}
\end{equation}
The pre-shock values of the Lorentz factor $\Gamma_\mathrm{fs,u}$ and the density $\rho_\mathrm{fs,u}$ of the ejecta are obtained as functions of the time $t$ by substituting the shock position into the profiles (\ref{eq:free_gamma}) and (\ref{eq:adiabatic_rho}),
\begin{equation}
\Gamma_\mathrm{fs,u}\simeq (m+1)^{1/2}\Gamma_\mathrm{fs},
\end{equation}
and
\begin{equation}
\rho_\mathrm{fs,u}\simeq \rho_0\left(\frac{t}{t_0}\right)^{-3}
(m+1)^{-n/2}\left(\frac{\Gamma_\mathrm{fs}}{\Gamma_\mathrm{max}}\right)^{-n},
\label{eq:preshock_rho}
\end{equation}
where we have assumed that the shock is ultra-relativistic, $\Gamma_\mathrm{fs}\gg 1$.

\subsubsection{Shock Jump Condition}
The post-shock Lorentz factor $\Gamma_\mathrm{fs,d}$ is found by solving Equation (\ref{app1:expression_for_gamma}). 
Introducing the ratio $y$ of the post-shock Lorentz factor to the shock Lorentz factor, 
\begin{equation}
\Gamma_\mathrm{fs,d}=y\Gamma_\mathrm{fs},
\end{equation}
one has to solve the following cubic equation to find $y$ (see Appendix \ref{jump_condition} for the derivation),
\begin{equation}
\gamma y^3+2(m+1)^{1/2}y^2-2y-\gamma(m+1)^{1/2}=0.
\end{equation}
The post-shock density and pressure are obtained from Equations (\ref{app1:expression_for_rho}) and (\ref{app1:expression_for_p}), 
\begin{equation}
\rho_\mathrm{fs,d}=\rho_\mathrm{fs,u}\frac{my}{(m+1)^{1/2}(y^2-1)}
\equiv f_{\rho}\rho_\mathrm{fs,u},
\end{equation}
and
\begin{equation}
p_\mathrm{fs,d}=\rho_\mathrm{fs,u}\frac{m(m+1-y^2)}{2(m+1)(y^2+m+1)}
\equiv f_p\rho_\mathrm{fs,u}. 
\end{equation}

It is worth noting that the time dependence of these quantities $\rho_\mathrm{fs,d}$ and $p_\mathrm{fs,d}$ are exactly same as that of the pre-shock density $\rho_\mathrm{fs,u}$. 
This is because both the pre-shock and the post-shock Lorentz factors are proportional to the shock Lorentz factor. 
Numerically evaluated values of $y$, $f_\rho$, and $f_p$ for some values of $m$ are presented in Table \ref{table:jump}.

\begin{figure}[tbp]
\begin{center}
\includegraphics[scale=0.45]{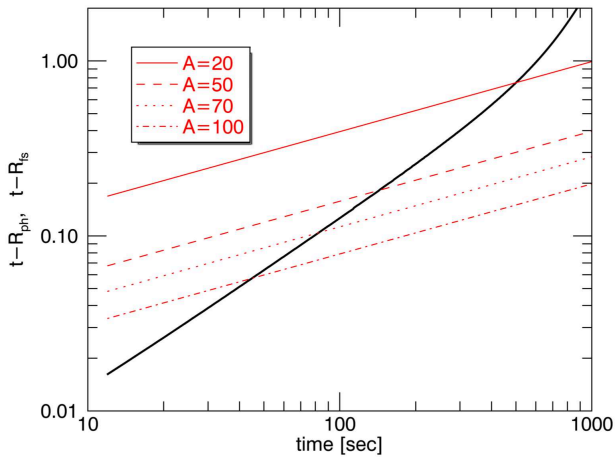}
\caption{
Temporal evolution of the positions of the photosphere and the forward shock for the ejecta model considered in Section \ref{sec:breakout_radius}. 
The distance of the photospheric radius (thick black line) and the forward shock (thin red lines) from $r=t$ is shown as functions of the time $t$. 
Forward shocks with $A=20$, $50$, $70$, and $100$ are presented.  
}
\label{fig:optical_depth}
\end{center}
\end{figure}

\subsection{Reverse Shock}
In the relativistic fireball, the reverse shock forms and converts the kinetic energy of the fireball into the internal energy of the shocked gas. 
When the reverse shock is at $r=R_\mathrm{rs}(t)$, the density and the pressure of the gas flowing into the shock front are expressed as follows,
\begin{equation}
\rho_\mathrm{rs,u}=\rho_\mathrm{in}\left(\frac{R_\mathrm{rs}}{R_\mathrm{in}}\right)^{-2},
\end{equation}
and
\begin{equation}
p_\mathrm{rs,u}=\frac{\rho_\mathrm{in}\epsilon_0}{3}\left(\frac{R_\mathrm{rs}}{R_\mathrm{in}}\right)^{-8/3}.
\end{equation}
Assuming a strong shock, one finds the post-shock density $\rho_\mathrm{rs,d}$ and the pressure $p_\mathrm{rs,d}$ from Equations (\ref{eq:jump_rho}) and (\ref{eq:jump_pre}),
\begin{equation}
\rho_\mathrm{rs,d}
=
\sqrt{2}
\frac{\eta}{\Gamma_\mathrm{rs}}
\rho_\mathrm{in}\left(\frac{R_\mathrm{rs}}{R_\mathrm{in}}\right)^{-2},
\end{equation}
and
\begin{equation}
p_\mathrm{rs,d}=\frac{2}{3}
\frac{\eta^2}{\Gamma_\mathrm{rs}^2}
\rho_\mathrm{in}\left(\frac{R_\mathrm{rs}}{R_\mathrm{in}}\right)^{-2}.
\end{equation}
From the thin shell approximation, the shock Lorentz factor evolves as,
\begin{equation}
\Gamma_\mathrm{rs}\propto t^{(1-\alpha)/3}.
\end{equation}
Therefore, the time dependence of these two variables is found to be,
\begin{equation}
\rho_\mathrm{rs,d}\propto t^{(\alpha-7)/3},
\end{equation}
and
\begin{equation}
p_\mathrm{rs,d}\propto t^{2(\alpha-4)/3}. 
\end{equation}
The time dependence of the pressure is identical with that derived under the thin shell approximation in Section \ref{thin_shell}. 

\begin{table}
\begin{center}
  \caption{Values of $y$, $f_\rho$, and $f_p$ for some values of the exponent $m$}
\begin{tabular}{cccc}
\hline\hline
m&$y$&$f_\rho$&$f_p$\\
\hline
-0.5& 0.952&7.18&0.107\\
-0.6& 0.937&7.31&0.191\\
-2/3& 0.926&7.45&0.282\\
\hline\hline
\end{tabular}
 \label{table:jump}
\end{center}
\end{table}

\subsection{Comparison with Numerical Simulation}
From the numerical simulation presented in the previous section, we find that the Lorentz factor $\Gamma_\mathrm{peak}$, the density $\rho_\mathrm{peak}$, and the pressure $p_\mathrm{peak}$ at the point where the density profile shows a peak evolve as, 
\begin{equation}
\Gamma_\mathrm{peak}\propto t^{0.32},\ \ \ 
\rho_\mathrm{peak}\propto t^{-1.8},\ \ \ 
p_\mathrm{peak}\propto t^{-2.5},
\end{equation}
by fitting a power-law function of the time $t$ from $t=50$ s to $t=100$ s. 
On the other hand, from the thin shell approximation and the theoretical consideration in this section, the temporal behavior of these variables with $\alpha=0$ should be
\begin{equation}
\Gamma_\mathrm{rs}\propto t^{0.33},\ \ \ 
\rho_\mathrm{rs,d}\propto t^{-2.3},\ \ \ 
p_\mathrm{rs,d}\propto t^{-2.7}.
\end{equation}
The exponents of the Lorentz factor and the pressure obtained from the numerical simulation are in good agreement with the theoretical values. 
The density decreases at a slower rate than the theoretical expectation. 
In fact, the density peak is located at the contact discontinuity separating the shocked fireball and the ejecta as seen in Figure \ref{fig:profile100}. 
At the contact discontinuity, the swept gas exhibits a sharp peak in the density and the temporal evolution of the peak value of the density seems to be significantly affected by the resolution of the numerical simulation. 
This is why the exponent of the temporal evolution of the peak density deviates from the theoretical value. 
On the other hand, the Lorentz factor and the pressure of the shocked fireball and the ejecta are continuous at the contact discontinuity. 
Therefore, the numerically obtained exponents of the Lorentz factor and the pressure well agree with the theoretical values. 

We also fit power-law functions of the time $t$ to the temporal evolution of the Lorentz factor, the density, and the pressure of the gas immediately behind the reverse shock from $t=65$ s to $t=100$ s and obtain the following scaling laws,
\begin{equation}
\Gamma_\mathrm{rs,sim}\propto t^{0.11},\ \ \ 
\rho_\mathrm{rs,sim}\propto t^{-2.2},\ \ \ 
p_\mathrm{rs,sim}\propto t^{-2.3}.
\end{equation}
While the exponent of the temporal evolution of the density agrees with the value expected from the analytical considerations, those of the Lorentz factor and the pressure deviate from the analytical values. 
For the forward shock, it is hard to correctly measure the values of the hydrodynamical variables behind the front because the shock structure gradually develops after the pressure of the ejecta decreases to a sufficiently small value as we have described in Section \ref{sec:structure_of_shell}.

\section{EXPECTED EMISSION FROM SHOCKED GAS}\label{sec:emission}
The ejecta are dense and opaque immediately after the formation and the photosphere is initially located at the outer edge of the ejecta. 
As the ejecta expand, the photosphere recedes from the forward shock. 
As a consequence, the photosphere eventually enters into the fireball. 
In this section, we summarize expected phenomena capable of producing bright X-ray or gamma-ray emission. 

\begin{figure}[tbp]
\begin{center}
\includegraphics[scale=0.5]{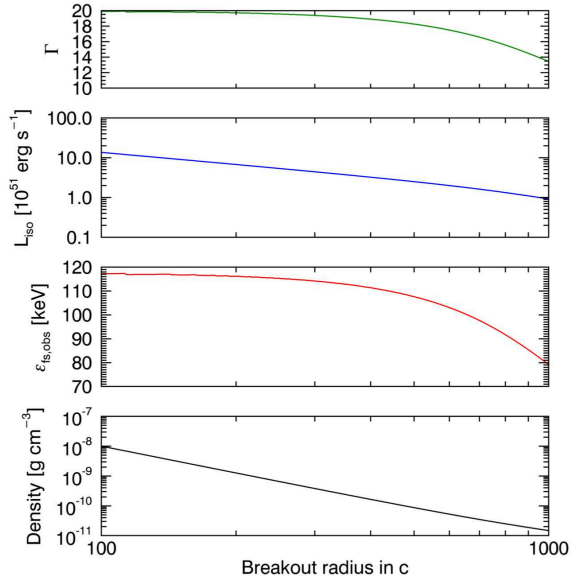}
\caption{
Lorentz factor, the estimated luminosity, the observed photon temperature, and the density at the photosphere as functions of the breakout radius. 
}
\label{fig:breakout}
\end{center}
\end{figure}

\subsection{Shock Breakout from Expanding Ejecta}\label{sec:breakout}
The forward shock propagating in the ejecta would emerge from the photosphere, which leads to bright shock breakout emission. 
From the analytical considerations, we find the ratio of the pre-shock Lorentz factor of the ejecta to the forward shock Lorentz factor to be $(1+m)^{-1/2}=1.7$. 
The corresponding relative velocity is $\sim 0.8$, suggesting a mildly relativistic shock breakout. 
There are a number of studies on the supernova shock breakout predicting a bright X-ray flash at the moment of the emergence of a radiative shock from the stellar atmosphere \citep{1974ApJ...187..333C,1978ApJ...223L.109K,1978ApJ...225L.133F}.
The stationary structure of a radiative shock in a radiation-dominated medium was investigated by several authors \citep[e.g.,][]{1976ApJS...32..233W,2010ApJ...716..781K,2010ApJ...725...63B}. 
Furthermore, \cite{2010ApJ...725..904N,2012ApJ...747...88N} studied the emission from the gas having been ejected from the stellar surface. 
The radiative shock emerging from the photosphere in a freely expanding gas might produce similar electromagnetic signals to those predicted by the earlier studies. 
In the following, we estimate the expected average photon energy and the isotropic luminosity of the flash by using a simplified model. 

\subsubsection{Breakout Radius}\label{sec:breakout_radius}
The photospheric radius of the ejecta is calculated as follows. 
The optical depth of the gas measured from a radius $r=r_\ast$ to the outer edge of the ejecta along the radial direction at $t=t_\ast$ is given by,
\begin{equation}
\tau(t_\ast,r_\ast)=\int_{r_\ast}^\infty\kappa\rho\Gamma(1-\beta)dr,
\end{equation}
 \citep[e.g.,][]{1991ApJ...369..175A}. 
We assume that the dominant opacity source is electron scattering, $\kappa=\kappa_\mathrm{es}=0.2$ cm$^{2}$ g$^{-1}$, which is reasonable for a fully ionized hot gas. 
It is worth noting that the density and the Lorentz factor of the gas evolve as the photon ray of interest moves toward the outer edge of the ejecta. 
We define the photospheric radius $R_\mathrm{ph}$ at $t=t_\ast$ as the radius where the thus calculated optical depth is equal to unity. 
Therefore, it is obtained by solving the following equation with respect to $R_\mathrm{ph}$,
\begin{equation}
\tau(t_\ast,R_\mathrm{ph})=1.
\end{equation}

We consider a freely expanding gas with the density, the pressure, and the Lorentz factor profiles given by Equations (\ref{eq:adiabatic_rho}), (\ref{eq:adiabatic_pre}), and (\ref{eq:free_gamma}). 
The parameters characterizing the profiles are $t_0=10$ s, $\Gamma_\mathrm{max}=20$, $\rho_0=10^{-5}$ g cm$^{-3}$, and $p_0/\rho_0\equiv f_0=0.03$. 
The exponents $n$ and $s$ are set to $n=1$ and $s=4n/3$, which gives a spatially uniform entropy profile, $p/\rho^{4/3}=\mathrm{Const.}$ 
The photospheric radius is numerically calculated for the given set of parameters and the distance $t-R_\mathrm{ph}$ between the photosphere and $r=t$ is shown as a function of the time $t$ in Figure \ref{fig:optical_depth}. 
On the other hand, specifying the value of $A$ in Equation (\ref{eq:gamma_fs}), the position of the forward shock is obtained from Equation (\ref{eq:shock_position}). 
The distance $t-R_\mathrm{fs}$ between the forward shock and $r=t$ is plotted in Figure \ref{fig:optical_depth} for $A=20$, $50$, $70$, and $100$. 
For a specific value of the parameter $A$, the breakout time when the forward shock emerges from the photosphere is obtained as the time satisfying  $R_\mathrm{ph}=R_\mathrm{fs}$, i.e., the intersection of the curves showing $t-R_\mathrm{ph}$ and $t-R_\mathrm{fs}$ in Figure \ref{fig:optical_depth}. 
The radius of the photosphere at the breakout time is called the breakout radius and denoted by $R_\mathrm{br}$. 
Larger values of the parameter $A$ indicate faster forward shocks, leading to earlier breakout times. 
This is why smaller breakout radii are realized for larger values of the parameter $A$ as shown in  Figure \ref{fig:optical_depth}. 

\subsubsection{Estimation of the Average Photon Energy in the Post-shock Gas}
Initially, the ejecta are sufficiently dense and equilibrium between radiation and matter is achieved. 
We assume that the internal energy of the ejecta is dominated by radiation. 
Denoting the post-shock pressure by $p_\mathrm{fs,d}$, the equilibrium photon temperature $T_\mathrm{eq}$ can be obtained by solving
\begin{equation}
a_\mathrm{r}T_\mathrm{eq}^4=3p_\mathrm{fs,d},
\end{equation}
where $a_\mathrm{r}$ is the radiation constant. 
On the other hand, the time $t_\mathrm{eq}$ required for the post-shock gas to achieve the equilibrium between matter and radiation by producing a sufficient number of photons via free-free process is estimated by dividing the internal energy density $a_\mathrm{r}T_\mathrm{eq}^4$ by the free-free emissivity $\epsilon_\mathrm{ff}$, 
\begin{equation}
t_\mathrm{eq}=\Gamma \frac{a_\mathrm{r}T_\mathrm{eq}^4}{\epsilon_\mathrm{ff}}
,
\end{equation}
where the Lorentz factor of the gas in the above expression is needed to convert the time scale in the comoving frame of the gas to that in the laboratory frame. 
This time scale is usually much longer at $t>100$ s than the elapsed time for parameters of interest. 
Therefore, we can assume that the photon production via free-free process after the passage of the forward shock is negligible. 
In such a situation, the internal energy produced by the dissipation of the shock kinetic energy is shared by ions, electrons, and photons swept by the shock. 
The ion and electron number density in the pre-shock gas with the density $\rho_\mathrm{fs,u}$ are estimated to be
\begin{equation}
n_\mathrm{ion}= \frac{n_\mathrm{e}}{Z_\mathrm{ion}}= \frac{\rho_\mathrm{fs,u}}{A_\mathrm{ion}m_\mathrm{u}},
\end{equation}
where the gas is assumed to be fully ionized and the mass and the atomic numbers of ions are denoted by $A_\mathrm{ion}$ and $Z_\mathrm{ion}$. 
The photon temperature $T_{\mathrm{fs,u}}$ and the photon number density $n_\mathrm{fs,u}$ in the pre-shock gas is estimated by 
\begin{equation}
T_{\mathrm{fs,u}}=\left(\frac{3p_\mathrm{fs,u}}{a_\mathrm{r}}\right)^{1/4},
\end{equation}
and 
\begin{equation}
n_\mathrm{fs,u}=\frac{a_\mathrm{r}T_\mathrm{fs,u}^3}{3k_\mathrm{B}},
\end{equation}
where $k_\mathrm{B}$ is the Boltzmann constant. 
For parameters of interest, the number density of photons is much larger than those of ions and electrons. 

These photons are tightly coupled with electrons through Compton scattering. 
Thus, at the shock front, the jump in the photon number density is same as that in the rest mass density. 
After the energy equipartition between gas and radiation is realized via Compton scattering, the average energy for a single photon would be
\begin{equation}
\epsilon_\mathrm{fs,d}=
\frac{3p_\mathrm{fs,d}}{f_{\rho}n_\mathrm{fs,u}},
\label{eq:average_photon_energy}
\end{equation}
when we assume that $n_\mathrm{fs,u}\gg n_\mathrm{e},n_\mathrm{ion}$. 
When the photosphere is present immediately after the forward shock, radiation with the doppler-boosted average photon energy,
\begin{equation}
\epsilon_\mathrm{fs,obs}=
\Gamma_\mathrm{fs,d}\epsilon_\mathrm{fs,d},
\label{eq:e_obs_fs}
\end{equation}
would be observed. 
The thus estimated observed photon energy is shown as a function of the breakout radius in Figure \ref{fig:breakout}

Although we have assumed that the number of  photons does not change in the course of the energy equipartition, some processes changing the number of photons and electrons, such as, double Compton scattering and pair production, may increase the number densities of photons and electrons and change the average photon energy. 
The thermal evolution of the mixture of gas and radiation toward the energy equipartition should be investigated in detail to find the accurate value of the average photon energy. 
In the following, we simply regard that emission with the photon energy given in Equation (\ref{eq:e_obs_fs}) is observed as the shock breakout emission. 

\subsubsection{Luminosity}
The isotropic luminosity of the breakout emission is estimated as follows. 
We have assumed that the internal energy of the post-shock gas is dominated by radiation. 
In addition, we assume that the radiation field in the post-shock gas is isotropic in the comoving frame of the gas and the gas is moving at ultra-relativistic speeds, $\Gamma_\mathrm{fs,d}\gg 1$.  
From the Lorentz transformation of the energy-momentum tensor of the radiation field, the radiative flux $F_\mathrm{fs}$ along the radial direction in the laboratory frame can be obtained as,
\begin{equation}
F_\mathrm{fs}\simeq \frac{4}{3}\Gamma_\mathrm{fs,d}^2u_\mathrm{fs,d}
\simeq 4\Gamma_\mathrm{fs,d}^2p_\mathrm{fs,d}.
\end{equation}
The isotropic luminosity for the breakout emission with the breakout radius $R_\mathrm{br}$ is estimated to be
\begin{equation}
L_\mathrm{fs}=16\pi f_\mathrm{fs}R_\mathrm{br}^2\Gamma_\mathrm{fs,d}^2p_\mathrm{fs,d},
\label{eq:Liso_break}
\end{equation}
where we have introduced a parameter $f_\mathrm{fs}$ representing the efficiency of the emission. 
The thus estimated luminosity with $f_\mathrm{fs}=1$ is shown as a function of the breakout radius in Figure \ref{fig:breakout}.

\subsection{Photospheric Emission from Reverse-Shocked Fireball}
After the shock breakout emission, the photospheric emission from the shocked fireball is expected. 
\subsubsection{Estimation of the Average Photon Energy in the Post-shock Gas}
We estimate the average photon energy of the post-shock gas in the same way as the breakout emission. 
We evaluate the photon number density immediately after the shock passage. 
The number density of photons is expected to be 
\begin{equation}
n_\mathrm{rs,u}=
\frac{a_\mathrm{r}T_\mathrm{ph}^3}{3k_\mathrm{B}}=
\frac{a_\mathrm{r}^{1/4}\rho_\mathrm{in}^{3/4}\epsilon_\mathrm{in}^{3/4}}{3k_\mathrm{B}}
\left(\frac{R_\mathrm{rs}}{R_\mathrm{in}}\right)^{-2}.
\end{equation}
before being swept by the shock. 
Thus, the post-shock value of the photon number density leads to
\begin{equation}
n_\mathrm{rs,d}=\sqrt{2}n_\mathrm{rs,u}\frac{\eta}{\Gamma_\mathrm{rs}}=
\frac{\sqrt{2}a_\mathrm{r}^{1/4}\rho_\mathrm{in}^{3/4}\epsilon_\mathrm{in}^{3/4}}{3k_\mathrm{B}}
\frac{\eta}{\Gamma_\mathrm{rs}}
\left(\frac{R_\mathrm{rs}}{R_\mathrm{in}}\right)^{-2}.
\end{equation}
We estimate the average photon energy $\epsilon_\mathrm{rs,d}$ by dividing the post-shock internal energy, which is given by $3p_\mathrm{rs,d}$, by the post-shock photon number density,
\begin{equation}
\epsilon_\mathrm{rs,d}
=
\frac{3\sqrt{2}k_\mathrm{B}\rho_\mathrm{in}^{1/4}}{a_\mathrm{r}^{1/4}\epsilon_\mathrm{in}^{3/4}}\frac{\eta}{\Gamma_\mathrm{rs}}
=
\frac{3\sqrt{2}k_\mathrm{B}L^{1/4}\eta^{3/4}}{(4\pi)^{1/4}a_\mathrm{r}^{1/4}\epsilon_\mathrm{in}^{3/4}R_\mathrm{in}^{1/2}\Gamma_\mathrm{in}^{1/4}\Gamma_\mathrm{rs}}.
\end{equation}
Since the post-shock Lorentz factor for the reverse shock with the Lorentz factor $\Gamma_\mathrm{rs}$ is given by
\begin{equation}
\Gamma_\mathrm{rs,d}=\sqrt{2}\Gamma_\mathrm{rs},
\end{equation}
the average photon energy in the observer frame leads to
\begin{equation}
\epsilon_\mathrm{rs,obs}=\Gamma_\mathrm{rs,d}\epsilon_\mathrm{rs,d}=
\frac{6k_\mathrm{B}L^{1/4}\eta^{3/4}}{(4\pi)^{1/4}a_\mathrm{r}^{1/4}\epsilon_\mathrm{in}^{3/4}R_\mathrm{in}^{1/2}\Gamma_\mathrm{in}^{1/4}}.
\end{equation}

\subsubsection{Luminosity}
Then, we estimate the isotropic luminosity of the photospheric emission when the photosphere is close to the reverse shock in the same way as the breakout emission. 
The isotropic luminosity $L_\mathrm{iso,rs}$ is estimated to be
\begin{eqnarray}
L_\mathrm{iso,rs}&\simeq& 16\pi f_\mathrm{rs}R_\mathrm{rs}^2\Gamma_\mathrm{rs}^2 p_\mathrm{rs,d}
\nonumber\\
&=&
\frac{32\pi}{3} f_\mathrm{rs}R_\mathrm{in}^2\eta^2\rho_\mathrm{in}=
\frac{8}{3} f_\mathrm{rs}L\frac{\eta}{\Gamma_\mathrm{in}},
\end{eqnarray}
where a parameter $f_\mathrm{rs}$ representing the efficiency of the emission has been introduced. 

\subsection{Implications to GRB Prompt Emission}
In the previous sections, we propose that the breakout emission from the forward shock in the ejecta and the photospheric emission from the reverse shocked fireball could contribute to the prompt emission of GRBs. 

As described in Section \ref{sec:breakout}, larger breakout radii are realized when smaller values of the parameter $A$ are assumed. 
The value of the parameter $A$ depends the energy and the mass of the ejecta and thus reflects the structure of the stellar envelope and the energy deposition from the fireball at the initial phase of the injection. 
Since a smaller value of $A$ represents a slower forward shock, the breakout occurs at a later phase of the dynamical evolution, resulting in the breakout in more dilute medium and less luminous emission. 
This is why a larger breakout radius (or equivalently smaller $A$) produce emission with smaller values of the average photon energy and the luminosity as shown in Figure \ref{fig:breakout}. 
Furthermore, for a fireball with a larger kinetic power and a mass injection rate, a more bright emission is expected. 

Recent observations of GRBs by the {\it Fermi} satellite have revealed temporal behaviors of the prompt gamma-ray emission in great detail. 
Especially, the delayed detection of GeV photons \citep{2009Sci...323.1688A,2010ApJ...712..558A} is one of the outstanding features of bursts observed by {\it Fermi}. 
In other words, spectra become harder in the later phase of the prompt emission.
In addition, components well fitted by Planck functions are found in the prompt emission in the first few seconds after the trigger \citep{2004ApJ...614..827R,2005ApJ...625L..95R,2006ApJ...652.1400R,2012ApJ...757L..31A}. 

From our model, the emission from the forward shock emerging from the photosphere in the expanding ejecta can be detected as an early electromagnetic signal. 
If we take a model with the breakout radius of $R_\mathrm{br}=200$ for example, the observed average photon energy and the isotropic luminosity are estimated to be 
\begin{eqnarray}
\hspace{-2.2em}\epsilon_\mathrm{fs,obs}&\sim& 120\ \mathrm{keV}
\left(\frac{\Gamma_\mathrm{fs}}{20}\right)
\left(\frac{f_\rho}{7.5}\right)^{-1}
\left(\frac{f_p}{0.28}\right)
\nonumber\\&&
\hspace{-3em}\times\left(\frac{\rho_\mathrm{fs,u}}{1.3\times 10^{-9}\ \mathrm{g}\ \mathrm{cm}^{-3}}\right)
\left(\frac{n_\mathrm{fs,u}}{1.4\times10^{19} \mathrm{cm}^{-3}}\right)^{-1},
\end{eqnarray}
and 
\begin{eqnarray}
L_\mathrm{fs}&\sim& 7\times 10^{51}f_\mathrm{fs}\ \mathrm{erg\ s}^{-1}
\left(\frac{R_\mathrm{br}}{6\times 10^{12}\ \mathrm{cm}}\right)^2
\nonumber\\&&
\times\left(\frac{\Gamma_\mathrm{fs}}{20}\right)^2
\left(\frac{f_p}{0.28}\right)
\left(\frac{\rho_\mathrm{fs,u}}{1.3\times 10^{-9}\ \mathrm{g}\ \mathrm{cm}^{-3}}\right)
,
\end{eqnarray}
which are in good agreement with the observed temperature and luminosity of the thermal components in BATSE bursts \citep{2004ApJ...614..827R,2005ApJ...625L..95R,2006ApJ...652.1400R}. 

After the shock emergence, the photosphere moves into the inner region of the shell and the photospheric emission from the reverse-shocked fireball starts contributing to the prompt gamma-ray emission. 
In this region, the kinetic power of the jet is converted to the internal energy of the shocked gas and escape as radiation. 
Thus, the luminosity of the emission from the reverse shock is constant as we have assumed the steady energy injection. 
For the fireball with $L=10^{51}$ erg s$^{-1}$, $\Gamma_\mathrm{in}=5$ and $\epsilon_\mathrm{in}=20$, which is corresponding to $\eta=138$, the observed average photon energy and the isotropic luminosity yield 
\begin{eqnarray}
\epsilon_\mathrm{rs,obs}&=&1\ \mathrm{MeV}
\left(\frac{L}{10^{51}\ \mathrm{erg\ s}^{-1}}\right)^{1/4}
\left(\frac{\eta}{138}\right)^{3/4}
\left(\frac{\epsilon_0}{20}\right)^{-3/4}
\nonumber\\&&
\times\left(\frac{R_\mathrm{in}}{10^9 \mathrm{cm}}\right)^{-1/2}
\left(\frac{\Gamma_\mathrm{in}}{5}\right)^{-1/4}
,
\end{eqnarray}
and 
\begin{eqnarray}
L_\mathrm{iso,rs}&=&7\times 10^{52}f_\mathrm{rs}\ \mathrm{erg\ s}^{-1}
\left(\frac{L}{10^{51}\ \mathrm{erg\ s}^{-1}}\right)
\nonumber\\&&
\times\left(\frac{\eta}{138}\right)
\left(\frac{\Gamma_\mathrm{in}}{5}\right)^{-1}
,
\end{eqnarray} 
These values are similar to the typical values of the spectral peak energy and the isotropic gamma-ray energy of GRBs. 
The shocked gas finally becomes transparent and the emission from the ultra-relativistic fireball can be seen. 
The delayed GeV emission might correspond to the emission from the ultra-relativistic fireball. 

The photospheric emission from the unshocked ejecta would also contribute to the prompt and the afterglow emission from GRBs in soft X-ray range. 
Recent discovery of a thermal component ($\sim0.1$-$1.0$ keV) in soft X-ray spectra of some bursts observed by {\it Swift} XRT \citep[see, e.g.,][]{2012MNRAS.427.2950S,2012MNRAS.427.2965S} has invoked discussions on the origin of the component. 
The hydrodynamical interaction between the ambient gas and the ejecta may be important in understanding the origin of the thermal X-ray emission as pointed out in \cite{2013ApJ...764L..12S}. 

\section{CONCLUSIONS AND DISUCSSIONS}\label{sec:conclusions}
In this paper, we have considered the hydrodynamical interaction of an ultra-relativistic fireball with a gas expanding almost freely and studied the dynamical evolution of the resultant geometrically thin shell in analytical and numerical ways. 
In the analytical considerations, the shell is assumed to have an infinitesimal width and the time dependence of the Lorentz factor is derived from the equation of motion of the shell. 
Then, we perform a simulation by using a one-dimensional special relativistic hydrodynamics code with AMR technique to resolve the inner structure of the shell. 
The resultant temporal evolution of the shell is compared with the analytical considerations. 

We point out a possibility that the emission from the forward and reverse shocks at the fireball-ejecta interface could contribute to the prompt gamma-ray emission of GRBs. 
Our findings indicate that the dynamical evolution of the gas ahead of the ultra-relativistic fireball is of critical importance in understanding the temporal behavior of the photospheric emission recently found in some bursts. 
We have estimated only the average photon energy and the isotropic luminosity expected in the breakout and the photospheric emission. 
To investigate the temporal evolutions of these quantities, detailed calculations on how the photosphere in the ejecta evolves with time are required. 
We regard investigations of the temporal evolution of the photospheric emission as a future work. 
We claim that it is needed to clarify whether the possibility proposed in this work is actually responsible for early emission from GRBs. 

Finally, we note some remarks on the present work. 
We create freely expanding ejecta by injecting a jet into a gas with a power-law density profile and then investigate the hydrodynamical interaction between the ejecta and the jet.
Freely expanding ejecta with different density structure might be realized in some short GRB progenitor. 
Although the analysis of the density structure of freely expanding ejecta resulting from a NS-NS merger in recent simulations of a NS-NS merger \citep{2013PhRvD..87b4001H,2014ApJ...784L..28N} revealed that the density profile is well described by a power-law function of the radius, we cannot exclude a possibility that ejecta with a more complex density profile could be created as a result of a NS-NS merger. 

While this work considers the dynamical evolution of the fireball and the ejecta in spherical symmetry, the gamma-ray emitting region of a GRB is thought to be highly collimated. 
The discrepancy between the spherical and jet models should be treated carefully. 
Earlier numerical studies of the jet propagation in a massive star \citep[e.g.][]{2003ApJ...586..356Z} revealed that the inner part of the jet is well described by the spherical fireball model. 
On the other hand, at earlier stages of the dynamical evolution of a GRB jet, when the jet propagates in the star, materials are accumulated on the head of the jet. 
After the jet emerges from the surface, the gas on the head of the jet expands in the lateral direction, which would lead to the ejecta with a mass smaller than that expected for spherical cases.
However, once the bulk Lorentz factor of the jet reaches to the critical value given by the inverse of the opening angle of the jet, $\Gamma\sim \theta_\mathrm{op}^{-1}$, the ejecta and the jet could be treated as a conical part of a spherical outflow until the Lorentz factor decreases to the critical value and the jet break occurs. 

\acknowledgments
Numerical calculations were in part carried out on the general-purpose PC farm at Center for Computational Astrophysics, National Astronomical Observatory of Japan. 
A.S. is supported by Grant-in-Aid for JSPS Fellows (26$\cdot$10618). 
This work is supported in part by the JSPS Grants-in-Aid for Scientific Research (23224004).

\appendix
\section{Derivation of Shock Jump condition for hydrodynamical variables}\label{jump_condition}
We describe the derivation of the shock jump condition at a strong shock propagating into a cold gas for the completeness of this paper. 
Details of the derivation can be found in some textbooks or review papers \citep[e.g.,][]{landau,MM2003}. 

The shock jump condition gives the relations between the physical variables of a gas in the upstream, $\rho_\mathrm{u}$, $\beta_\mathrm{u}$, and $p_\mathrm{u}$, and those in the downstream, $\rho_\mathrm{d}$, $\beta_\mathrm{d}$, and $p_\mathrm{d}$. 
The corresponding Lorentz factors are $\Gamma_\mathrm{u}$ and $\Gamma_\mathrm{d}$ for flows in the upstream and downstream. 
We assume that the pressure of the gas in the upstream is negligible, $p_\mathrm{u}\ll \rho_\mathrm{u}$, and the flow is highly relativistic, $\Gamma_\mathrm{u},\Gamma_\mathrm{d}\gg 1$. 
From hydrodynamical equations for one-dimensional plane-parallel flows, one finds the following relations for the physical variables of the gas in the upstream and the downstream of a shock propagating at a velocity of $\beta_\mathrm{s}$ (the corresponding Lorentz factor is denoted by $\Gamma_\mathrm{s}$), the mass conservation,
\begin{equation}
(\rho_\mathrm{u}\Gamma_\mathrm{u}-\rho_\mathrm{d}\Gamma_\mathrm{d})\beta_\mathrm{s}
=
\rho_\mathrm{u}\Gamma_\mathrm{u}\beta_\mathrm{u}
-\rho_\mathrm{d}\Gamma_\mathrm{d}\beta_\mathrm{d},
\end{equation}
the momentum conservation,
\begin{equation}
(\rho_\mathrm{u}\Gamma_\mathrm{u}^2\beta_\mathrm{u}-\rho_\mathrm{d}h_\mathrm{d}\Gamma_\mathrm{d}^2\beta_\mathrm{d})\beta_\mathrm{s}
=
\rho_\mathrm{u}\Gamma_\mathrm{u}^2\beta_\mathrm{u}^2
-(\rho_\mathrm{d}h_\mathrm{d}\Gamma_\mathrm{d}^2\beta_\mathrm{d}^2+p_\mathrm{d}),
\end{equation}
and the energy conservation,
\begin{equation}
(\rho_\mathrm{u}\Gamma_\mathrm{u}^2-\rho_\mathrm{d}h_\mathrm{d}\Gamma_\mathrm{d}^2+p_\mathrm{d})\beta_\mathrm{s}
=
\rho_\mathrm{u}\Gamma_\mathrm{u}^2\beta_\mathrm{u}
-\rho_\mathrm{d}h_\mathrm{d}\Gamma_\mathrm{d}^2\beta_\mathrm{d},
\end{equation}
with
\begin{equation}
h_\mathrm{d}=1+\frac{{\gamma}}{{\gamma}-1}\frac{p_\mathrm{d}}{\rho_\mathrm{d}},
\end{equation}
where $h_\mathrm{d}$ and $\gamma$ are the specific enthalpy and the adiabatic index of the gas in the downstream. 
These equations can be rewritten as follows,
\begin{eqnarray}
\rho_\mathrm{u}\Gamma_\mathrm{u}(\beta_\mathrm{u}-\beta_\mathrm{s})
&=&
\rho_\mathrm{d}\Gamma_\mathrm{d}(\beta_\mathrm{d}-\beta_\mathrm{s}),
\label{app:mass}\\
\rho_\mathrm{u}\Gamma_\mathrm{u}^2\beta_\mathrm{u}(\beta_\mathrm{u}-\beta_\mathrm{s})
&=&
\rho_\mathrm{d}h_\mathrm{d}\Gamma_\mathrm{d}^2\beta_\mathrm{d}(\beta_\mathrm{d}-\beta_\mathrm{s})+p_\mathrm{d},
\label{app:momentum}\\
\rho_\mathrm{u}\Gamma_\mathrm{u}^2(\beta_\mathrm{u}-\beta_\mathrm{s})
&=&
\rho_\mathrm{d}h_\mathrm{d}\Gamma_\mathrm{d}^2(\beta_\mathrm{d}-\beta_\mathrm{s})+p_\mathrm{d}\beta_\mathrm{s},
\label{app:energy}
\end{eqnarray}
and some algebraic manipulations in the above expressions lead to the following equation,
\begin{equation}
\frac{\gamma}{\gamma-1}\Gamma_\mathrm{u}(\beta_\mathrm{u}-\beta_\mathrm{d})\Gamma_\mathrm{d}^2(\beta_\mathrm{d}-\beta_\mathrm{s})
=
\Gamma_\mathrm{u}(1-\beta_\mathrm{u}\beta_\mathrm{s})
-\Gamma_\mathrm{d}(1-\beta_\mathrm{d}\beta_\mathrm{s}).
\end{equation}
Since the flows are highly relativistic,
$\Gamma_\mathrm{u},\Gamma_\mathrm{d},\Gamma_\mathrm{s}\gg 1$, 
one obtains the following approximated expression of the above equation,
\begin{equation}
\frac{\Gamma_\mathrm{d}^2}{\Gamma_\mathrm{s}^2}=
\frac{\gamma\Gamma_\mathrm{u}+(2-\gamma)\Gamma_\mathrm{d}}
{(2-\gamma)\Gamma_\mathrm{u}+\gamma\Gamma_\mathrm{d}}.
\label{app1:expression_for_gamma}
\end{equation}
One can find the Lorentz factor $\Gamma_\mathrm{d}$ of the flow in the downstream for a given set of the Lorentz factor of the gas in the upstream and the shock Lorentz factor, $\Gamma_\mathrm{u}$ and $\Gamma_\mathrm{s}$, by solving this equation. 

Equation (\ref{app:mass}) can be solved for the density $\rho_\mathrm{d}$ of the flow in the downstream and approximated under the assumption of highly relativistic flows as follows,
\begin{equation}
\rho_\mathrm{d}=\rho_\mathrm{u}
\frac{\Gamma_\mathrm{d}(\Gamma_\mathrm{u}^2-\Gamma_\mathrm{s}^2)}
{\Gamma_\mathrm{u}(\Gamma_\mathrm{d}^2-\Gamma_\mathrm{s}^2)}.
\label{app1:expression_for_rho}
\end{equation}
Furthermore, the elimination of the enthalpy $h_\mathrm{d}$ from Equations (\ref{app:momentum}) and (\ref{app:energy}) yields
\begin{equation}
p_\mathrm{d}=\frac{\rho_\mathrm{u}\Gamma_\mathrm{u}^2
(\beta_\mathrm{u}-\beta_\mathrm{d})(\beta_\mathrm{u}-\beta_\mathrm{s})}
{1-\beta_\mathrm{d}\beta_\mathrm{s}},
\end{equation}
which is approximated as,
\begin{equation}
p_\mathrm{d}=\frac{
\rho_\mathrm{u}(\Gamma_\mathrm{u}^2-\Gamma_\mathrm{d}^2)
(\Gamma_\mathrm{u}^2-\Gamma_\mathrm{s}^2)
}
{2\Gamma_\mathrm{u}^2(\Gamma_\mathrm{d}^2+\Gamma_\mathrm{s}^2)}.
\label{app1:expression_for_p}
\end{equation}
Therefore, one finds the density and the pressure of the gas in the downstream from Equations (\ref{app1:expression_for_rho}) and (\ref{app1:expression_for_p}), once the Lorentz factor $\Gamma_\mathrm{d}$ of the flow in the downstream is obtained. 

Here we consider a special case with $\Gamma_\mathrm{u}\gg \Gamma_\mathrm{d},\Gamma_\mathrm{s}$, which corresponds to the reverse shock propagating in the fireball in this study. 
In this limit, Equation (\ref{app1:expression_for_gamma}) can be solved analytically,
\begin{equation}
\Gamma_\mathrm{d}=
\left(\frac{{\gamma}}{2-{\gamma}}\right)^{1/2}\Gamma_\mathrm{s}.
\end{equation}
Furthermore, the rest mass density and the pressure in the downstream are found to be,
\begin{equation}
\rho_\mathrm{d}
=\frac{{\gamma}}{2({\gamma}-1)}
\rho_\mathrm{u}\frac{\Gamma_\mathrm{u}}{\Gamma_\mathrm{d}}
=
\frac{{\gamma}}{2({\gamma}-1)}
\left(\frac{{\gamma}}{2-{\gamma}}\right)^{-1/2}
\rho_\mathrm{u}\frac{\Gamma_\mathrm{u}}{\Gamma_\mathrm{s}},
\label{eq:jump_rho}
\end{equation}
and
\begin{equation}
p_\mathrm{d}=(2-\gamma)
\rho_\mathrm{u}\frac{\Gamma_\mathrm{u}^2}{\Gamma_\mathrm{s}^2}.
\label{eq:jump_pre}
\end{equation}

\section{Numerical Techniques}\label{numerical_techniques}
In this section, we briefly describe our method to numerically integrate hydrodynamical equations. 

Equations (\ref{eq:continuity}) - (\ref{eq:enthalpy}) are numerically integrated by using a standard finite-volume method, i.e., the hydrodynamical variables averaged over each cell are evolved. 
We use the 3rd-order MUSCL scheme to obtain the values at the surfaces of the cell and then the numerical fluxes are calculated by the relativistic HLLC scheme \citep{2005MNRAS.364..126M}.

\subsection{Adaptive Mesh Refinement Technique}
The adaptive mesh refinement (AMR) technique \citep{1989JCoPh..82...64B} is now commonly used in various codes for astrophysical simulations, including some publicly available codes for hydrodynamics, such as, FLASH \citep{2000ApJS..131..273F}, ENZO \citep{2014ApJS..211...19B}, and so on. 
The implementation of the AMR technique in our code is realized by the well-known block-structured mesh technique. 

The whole numerical domain is covered by so-called AMR blocks. 
A unit AMR block is composed of 8 cells covering a part of the whole numerical domain and a few cells for the communications with other blocks. 
If some conditions (referred to as the "refinement criteria") are satisfied for a block and the level of the block is lower than the maximum refinement level, two other blocks with finer resolution, which is called "child blocks", are created and they cover the original block (referred to as the "parent block"). 
On the newly created blocks, the physical variables are interpolated from the parent block. 
The code calculates the temporal evolution of physical variables averaged over a cell. 
The volume average of a variable $A$ over $i$the cell is written as follows,
\begin{equation}
A_i=\frac{1}{V_i}\int AdV,
\end{equation}
where $V_i$ denotes the volume of the cell and the volume integral runs over the cell. 
When a couple of new blocks are created, the physical variables are interpolated from the parent block so that the volume-integrated value of the variable is conserved, 
\begin{equation}
A_i^\mathrm{p}V_i^\mathrm{p}= A_j^\mathrm{c}V^\mathrm{c}_j+A_{j+1}^\mathrm{c}V^\mathrm{c}_{j+1}.
\end{equation}
Here $A_i^\mathrm{p}$ and $V_i^\mathrm{p}$ are the volume-averaged variable in $i$the cell and the volume of the cell in the parent block. 
The $j$th and $(j+1)$th cells in the child block are assumed to be covered by the $i$th cell in the parent block and $A_j^\mathrm{c}$ and $V_j^\mathrm{c}$ are the physical variable and the volume corresponding to the $j$th cell in the child block. 
The physical variables in the newly created blocks are evolved according to the hydrodynamical equations with appropriate boundary conditions. 
Various refinement criteria can be used depending on the purpose of simulations. 
On the other hand, if a region is covered by cells with unnecessarily fine resolution, the resolution is coarsened by discarding some blocks. 
The synchronized time step is adopted in the current version of the code, i.e., the time step is same for all levels. 

\begin{figure}[tbp]
\begin{center}
\includegraphics[scale=0.5]{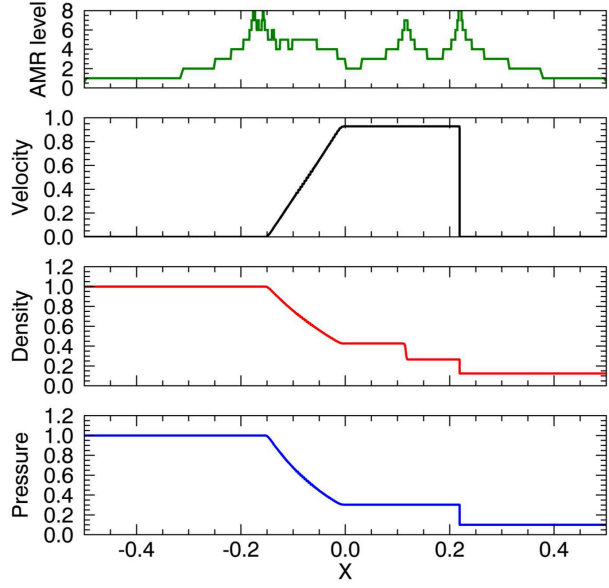}
\caption{Results of the Sod's shock tube test with the maximum refinement level of $8$. The panels represent the refinement level, velocity, density, and pressure profiles from top to bottom.}
\label{fig:sod}
\end{center}
\end{figure}

\begin{figure}[tbp]
\begin{center}
\includegraphics[scale=0.7]{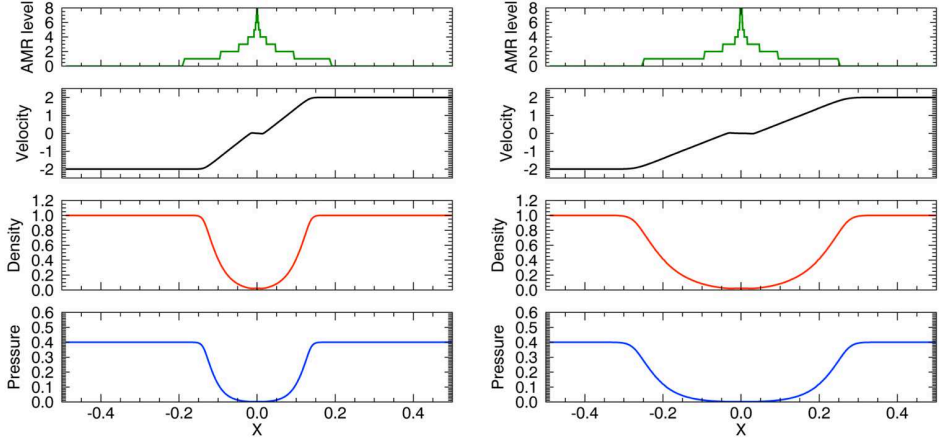}
\caption{Results of the Einfeldt's strong rarefaction test (1-2-0-3) with the maximum refinement level of $8$. Left and right panels correspond to the profiles of some physical variables at $t=0.05$ and $0.1$. }
\label{fig:ein}
\end{center}
\end{figure}

\subsection{Some One-dimensional Test Problems}
We carry out calculations of the following test problems to confirm that the developed code works well. 
\subsubsection{Sod's Shock Tube Test}
In this test problem, a domain, $-1\leq x\leq 1$, is initially separated into the following two states,
\begin{equation}
(\rho,v,p)=
\left\{\begin{array}{ccl}
(1.0,0.0,1.0)&\mathrm{for}&x\leq 0.0,\\
(0.125,0.0,0.1)&\mathrm{for}&0.0<x.
\end{array}\right.
\end{equation}
The domain is covered by 256 cells with the refinement level of $l=0$ and the maximum refinement level is set to $l_\mathrm{max}=8$. 
After the simulation starts, a shock wave and a rarefaction wave form and start propagating into the $+x$- and $-x$-directions. 
The gas is separated by the contact discontinuity, where the velocity and the pressure are continuous while the density shows a jump. 

The resultant profiles of the velocity, the density, and the pressure are shown in Figure \ref{fig:sod} and agree with the exact solution. 
The AMR level is also presented in the top panel of Figure \ref{fig:sod}. 
The shock front, the contact discontinuity, and the rarefaction front are well resolved. 

\subsubsection{Einfeldt's Strong Rarefaction Test}
We also carry out a test problem known as Einfeldt's 1-2-0-3 problem. 
In this problem, the computational domain is divided into to the following two states,
\begin{equation}
(\rho,v,p)=
\left\{\begin{array}{ccl}
(1.0,-2.0,0.4)&\mathrm{for}&x\leq 0.0,\\
(1.0,2.0,0.4)&\mathrm{for}&0.0<x.
\end{array}\right.
\end{equation}
The number of cells covering the domain and the maximum refinement level are same as the previous test problem. 
Snapshots of the physical variables for the test problem at $t=0.05$ and $0.1$ are shown in Figure \ref{fig:ein}. 
As the initial state contains a sharp discontinuity in the velocity at $x=0$, the computational domain around $x=0$ is covered by blocks with higher resolution. 
After the simulation starts, the discontinuity breaks up into a couple of rarefaction waves, which propagate into $\pm x$-directions. 

\subsubsection{Sedov-Taylor Point Explosion Test}

\begin{figure}[tbp]
\begin{center}
\includegraphics[scale=0.5]{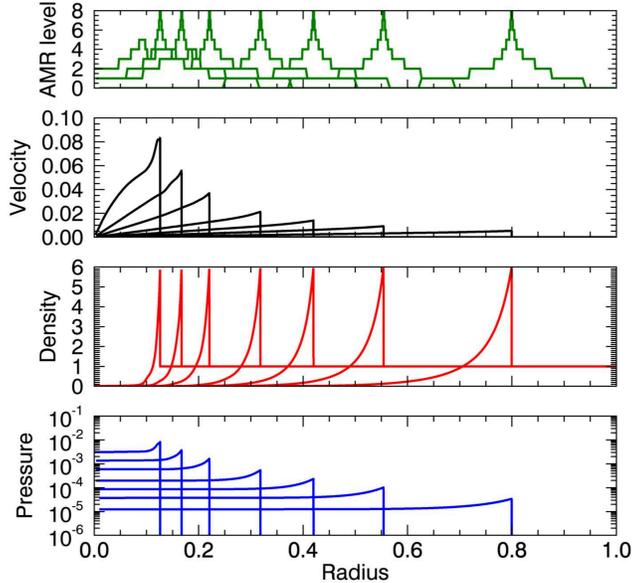}
\caption{Results of the Sedov-Taylor point explosion test with the maximum refinement level of $8$. The panels represent the refinement level, velocity, density, and pressure profiles from top to bottom.}
\label{fig:sedov}
\end{center}
\end{figure}

The Sedov-Taylor point explosion is a well-known problem of non-relativistic hydrodynamics in spherical symmetry \citep[see, e.g.,][]{1959sdmm.book.....S,1967pswh.book.....Z}. 
Initially, a region with high pressure (thermal bomb) is set in a small region surrounded by a cold and uniform medium. 
Then, a blast wave forms and propagates in the surrounding medium. 
Under a strong shock approximation, it is known that the profiles of hydrodynamical variables can be described by a self-similar solution after effects of the initial condition disappear. 

In this test problem, we set the computational domain to be $r\in [0,1]$ and assume a static and uniform medium,
\begin{equation}
\rho=1.0,\ \ \ \mathrm{and}\ \ \ v=0.0.
\end{equation}
The formation of a strong shock wave is realized by imposing the following initial condition for the pressure,
\begin{equation}
p=\epsilon_{p}+(1-\epsilon_{p})\exp(-r^2/r_0^2),
\end{equation}
with $r_0=0.02$ and $\epsilon_{p}=10^{-8}$. 
The adiabatic index of the gas is set to $\gamma=7/5$ in this problem. 
The computational domain is divided into 16 AMR blocks at the coarsest level ($l=0$). 
Thus, the domain is covered by $8\times 16=128$ cells at level $0$. 
The maximum refinement level is set to $l_\mathrm{max}=8$. 

Snapshots of the radial profiles of the velocity, the density, and the pressure are shown in Figure \ref{fig:sedov}. 
In the top panel of Figure \ref{fig:sedov}, the refinement level is also plotted. 
The shock front is covered by cells with the finest resolution. 
The expected density jump is $(\gamma+1)/(\gamma-1)=6$ under the strong shock approximation. 
The density profiles in Figure \ref{fig:sedov} show that the shock front is successfully resolved by the AMR technique. 
The profiles of the hydrodynamical variables eventually show the self-similarity and well agree with the exact self-similar solution.

\subsubsection{Special Relativistic Shock Tube Test}
\begin{figure}[tbp]
\begin{center}
\includegraphics[scale=0.8]{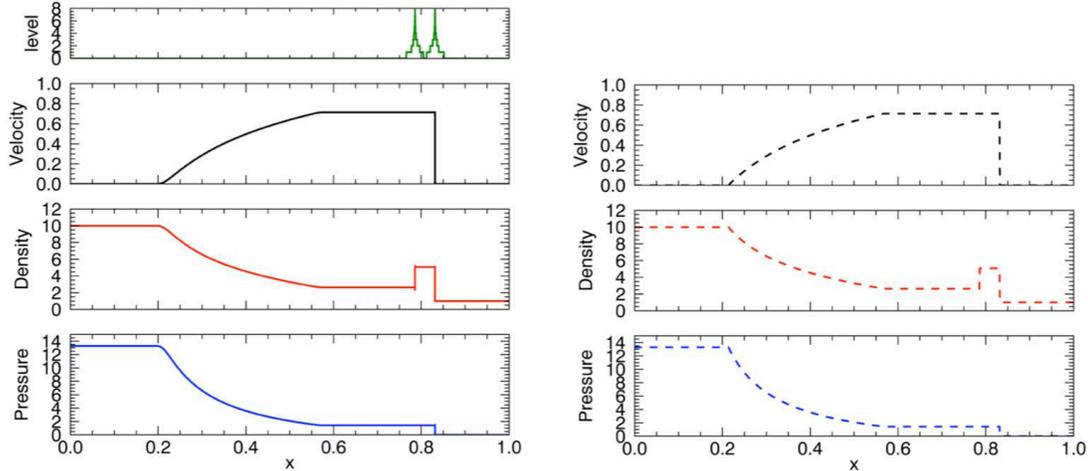}
\caption{Results of the relativistic shock tube test with the maximum refinement level of $8$. 
The profiles of the refinement level, velocity, density and the pressure are presented in the left panel. 
The exact solution of the problem is shown in the right panel. }
\label{fig:sr_shock_tube}
\end{center}
\end{figure}

This test is an extension of the Sod's shock tube test in special relativistic hydrodynamics. 
Initially, a domain, $0\leq x\leq 1$, is separated into the following two states,
\begin{equation}
(\rho,v,p)=
\left\{\begin{array}{ccl}
(10.0,0.0,13.3)&\mathrm{for}&x\leq 0.5,\\
(1.0,0.0,10^{-5})&\mathrm{for}&0.5<x.
\end{array}\right.
\end{equation}
Resultant profiles of the velocity, the density, and the pressure at $t=0.4$ are shown in the left panel of Figure \ref{fig:sr_shock_tube}. 
The refinement level is plotted as a function of the coordinate $x$ in the left panel of Figure \ref{fig:sr_shock_tube}. 
The shock wave propagating into the right boundary and the contact discontinuity are covered by cells with the finest resolution. 
In the right panel, the exact solution of the problem is shown. 
The profiles calculated by our code are in good agreement with the exact solution.

\end{document}